\documentclass{article}

\usepackage[utf8]{inputenc}
\usepackage[T1]{fontenc}
\usepackage{microtype}
\usepackage{amsmath,amssymb}
\usepackage{graphicx}
\usepackage{booktabs}
\usepackage{tikz}
\usetikzlibrary{arrows.meta}
\usepackage{xcolor}
\PassOptionsToPackage{hyphens}{url}
\usepackage{hyperref}

\usepackage{mlsys2024}
\gdef\isaccepted{1}
\makeatletter
\renewcommand{\Notice@String}{Preprint.}
\makeatother

\mlsystitlerunning{MatrixFSDP: Communication-Free Matrix Optimizers under ZeRO-3}

\newcommand{\sys}{MatrixFSDP}
\newcommand{\NS}{\mathrm{NS}}

\begin{document}

\twocolumn[
\mlsystitle{MatrixFSDP: Communication-Free Matrix Optimizers under\\ ZeRO-3 Parameter Sharding}

\begin{mlsysauthorlist}
\mlsysauthor{Ming Gao}{upitt}
\mlsysauthor{Yanwu Xu}{google}
\mlsysauthor{Hao Zhang}{tsinghua}
\end{mlsysauthorlist}
\mlsysaffiliation{upitt}{University of Pittsburgh}
\mlsysaffiliation{google}{Google}
\mlsysaffiliation{tsinghua}{Tsinghua University}
\mlsyscorrespondingauthor{Ming Gao}{dujinshidai30@gmail.com}
\mlsyscorrespondingauthor{Yanwu Xu}{yvansxu@gmail.com}
\mlsyscorrespondingauthor{Hao Zhang}{zhanghao\_990127@163.com}
\mlsyskeywords{distributed training, Muon, FSDP, ZeRO-3, optimizer-aware sharding}

\vskip 0.3in

\begin{abstract}
Matrix optimizers such as Muon can improve convergence and token efficiency by coupling coordinates
within each weight matrix. Muon orthogonalizes momentum-smoothed updates with Newton--Schulz and
therefore requires each complete 2D matrix as optimizer input. This conflicts with FSDP/ZeRO-3, whose
memory savings come from exposing only a shard of each parameter and gradient to every rank. Existing
ZeRO-3 integrations must therefore communicate matrix-wide information during the optimizer step,
either by reconstructing each matrix or by distributing its coupled matrix operations. ZeRO-1 owner
placement makes the update local, but keeps full parameters resident. We present \sys{}, which makes
whole-matrix ownership a ZeRO-3 shard layout. Each
data-parallel-sharded matrix is resident in full on one owner rank and empty elsewhere; the ZeRO-3
backward reduction consequently leaves the complete gradient on that owner, where Newton--Schulz runs
without an optimizer-step matrix collective. Across the shard group, parameters still have only one
resident copy. Making this uneven placement practical requires global owner planning with block-local
execution, communication that moves only nonempty owner segments, and explicit management of FSDP2
buffers and checkpoint shards. \sys{} provides these mechanisms as one end-to-end FSDP2 runtime and
supports Muon, Shampoo, and SOAP without distributed optimizer variants. Across 8--64 A100s, it
reduces optimizer-step latency over stock FSDP2-Muon by $4.2\times$--$54.6\times$, reaches up to
$3.3\times$ end-to-end speedup, and preserves ZeRO-3-scale memory for model sizes at which ZeRO-1
owner placement exceeds an 80\,GB GPU.
\par\noindent
Code is available at \url{https://github.com/findlamp/matrixfsdp-official}.
\end{abstract}
]

\printAffiliationsAndNotice{}

\section{Introduction}

Large-model training efficiency depends on both systems throughput and optimization efficiency: a
faster step is useful only together with a competitive number of steps and tokens to reach a target
quality. Matrix-structured optimizers improve the latter by exploiting interactions within each weight
matrix. Muon~\cite{muon}, for example, orthogonalizes momentum-smoothed updates and has demonstrated
strong token efficiency in large-scale training~\cite{moonshot,essentialai};
Shampoo~\cite{shampoo,anil2020} and SOAP~\cite{soap} use related matrix preconditioners. Their matrix
operations, however, require a complete 2D optimizer input. An arbitrary row-band does not produce the
same update as the full matrix.

That requirement conflicts with the interface that makes FSDP2/ZeRO-3~\cite{zero,fsdp} memory
efficient. After backward, each rank holds only slices of the parameters, gradients, and optimizer
state. Coordinate-wise AdamW can update those slices directly; Muon must communicate across them,
either by reconstructing the matrix or by running distributed Newton--Schulz. Existing systems
therefore occupy two endpoints. ZeRO-3 integrations retain sharded memory but add matrix-wide
communication to the post-backward critical path. ZeRO-1 owner systems make each matrix update local,
but retain full parameters on every rank. Neither provides local whole-matrix updates with ZeRO-3
parameter memory.

To bridge these endpoints, we present \sys{}, an FSDP2 runtime that changes the placement of ZeRO-3
shards rather than the optimizer being computed. Each data-parallel-sharded 2D weight is resident in
full on one owner rank and empty on the others; non-matrix tensors are packed into owner roles and
remain on AdamW. The existing backward reduction then deposits the complete matrix gradient on its
owner, eliminating optimizer-step matrix reconstruction while retaining one resident parameter copy
across the shard group.

This placement moves the systems problem into the FSDP2 runtime. Whole-matrix owners create uneven
resident memory and optimizer work; their shards no longer fit equal-size collectives and can create
fanout bottlenecks; and FSDP2 must safely reuse transient full-parameter storage across forward,
reshard, backward, optimization, and checkpointing. \sys{} addresses these coupled problems while
keeping materialization at FSDP-block granularity, so communication and computation retain FSDP2's
prefetch and overlap opportunities. Our contributions are:

\noindent$\triangleright$ \textbf{Owner-shaped ZeRO-3.} We formulate whole-matrix ownership as a
ZeRO-3 placement. Backward delivers each full 2D gradient directly to one owner, making Muon local and
the optimizer step communication-free without replicating parameters. The current design applies to
data-parallel-sharded matrices; tensor-parallel-owned fragments remain outside its scope.

\noindent$\triangleright$ \textbf{A runtime for uneven owners.} A global balance-aware planner assigns
matrix and tail roles across the model, while block-local execution preserves FSDP2 prefetch and
overlap. \texttt{MatrixShard} metadata and deterministic owner-segment P2P communication move only
nonempty shards, avoiding largest-owner padding and full-matrix all-gathers.

\noindent$\triangleright$ \textbf{End-to-end integration and evaluation.} Owner-buffer pinning,
mixed Muon/AdamW routing, and owner-shard checkpoint resharding carry the same placement through a
complete FSDP2 training path. Across 8--64 A100s, \sys{} preserves full-matrix optimizer semantics and
ZeRO-3-scale memory, provides up to $3.3\times$ end-to-end speedup, and supports Muon, Shampoo, and
SOAP without optimizer-step matrix collectives.

\section{Motivation}\label{sec:motivation}

In FSDP2 / ZeRO-3~\cite{zero,fsdp},
each rank stores only a slice of each parameter. Full weights exist transiently for forward/backward,
then are immediately resharded; after backward, each rank again holds only its local parameter and
gradient slices. This is exactly enough for AdamW, whose update is coordinate-wise. Once the gradient
slice is reduced, the optimizer step needs no further model-sized collective.

Muon~\cite{muon} changes that contract. For a 2D weight
with momentum-buffered gradient $G$, Muon applies a Newton--Schulz orthogonalization whose steps are
built from products such as $XX^\top$. Those products couple the rows of $G$: a row-band is not a
valid input to the same update. Shampoo~\cite{shampoo,anil2020} and SOAP~\cite{soap} have the same
shape of problem: their preconditioners are also whole-matrix operators. Under ordinary equal ZeRO-3
sharding, the runtime must therefore either reconstruct each 2D gradient or distribute the coupled
matrix algebra; local slice-wise optimization is not the same algorithm.

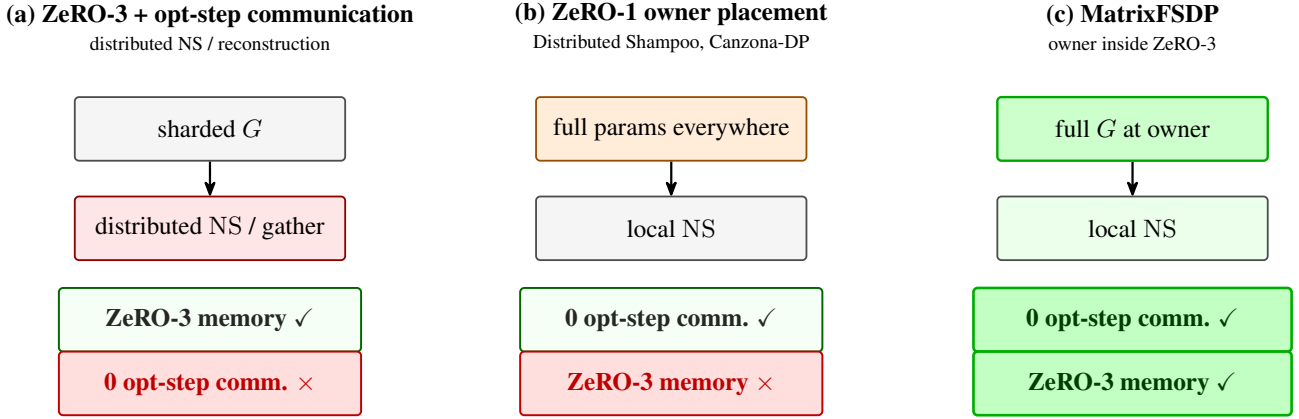
\begin{figure*}[t]
\centering
\resizebox{\textwidth}{!}{%
\begin{tikzpicture}[
  font=\small, >={Stealth[round]}, thick,
  title/.style={align=center, font=\small},
  stage/.style={draw=black!70, rounded corners=1.5pt, line width=0.65pt,
    minimum height=8mm, minimum width=34mm, align=center, font=\footnotesize},
  badge/.style={draw, rounded corners=1.5pt, line width=0.7pt,
    minimum height=8mm, minimum width=38mm, align=center, font=\footnotesize\bfseries},
  mutedgood/.style={badge, fill=green!4, draw=green!40!black, text=black!85},
  bad/.style={badge, fill=red!13, draw=red!75!black, text=red!75!black},
  stronggood/.style={badge, fill=green!24, draw=green!68!black, text=green!24!black,
    line width=0.95pt, minimum width=40mm, font=\footnotesize\bfseries},
]
\def\cA{0}\def\cB{5.8}\def\cC{11.6}
\node[title] at (\cA,2.62) {\textbf{(a) ZeRO-3 + opt-step communication}\\{\scriptsize distributed NS / reconstruction}};
\node[title] at (\cB,2.62) {\textbf{(b) ZeRO-1 owner placement}\\{\scriptsize Distributed Shampoo, Canzona-DP}};
\node[title] at (\cC,2.62) {\textbf{(c) \sys{}}\\{\scriptsize owner inside ZeRO-3}};
\node[stage, fill=gray!8] (a1) at (\cA,1.35) {sharded $G$};
\node[stage, fill=red!10, draw=red!55!black] (a2) at (\cA,0.10) {distributed $\NS$ / gather};
\draw[->] (a1) -- (a2);
\node[mutedgood] at (\cA,-1.05) {ZeRO-3 memory \checkmark};
\node[bad] at (\cA,-1.85) {0 opt-step comm. $\times$};
\node[stage, fill=orange!15, draw=orange!60!black] (b1) at (\cB,1.35) {full params everywhere};
\node[stage, fill=gray!8] (b2) at (\cB,0.10) {local $\NS$};
\draw[->] (b1) -- (b2);
\node[mutedgood] at (\cB,-1.05) {0 opt-step comm. \checkmark};
\node[bad] at (\cB,-1.85) {ZeRO-3 memory $\times$};
\node[stage, fill=green!20, draw=green!65!black, line width=0.85pt] (c1) at (\cC,1.35) {full $G$ at owner};
\node[stage, fill=green!8, draw=black!70] (c2) at (\cC,0.10) {local $\NS$};
\draw[->] (c1) -- (c2);
\node[stronggood] at (\cC,-1.05) {0 opt-step comm. \checkmark};
\node[stronggood] at (\cC,-1.85) {ZeRO-3 memory \checkmark};
\end{tikzpicture}%
}
\caption{Three ways to combine sharding with a whole-matrix optimizer. \textbf{(a)} Keeping ordinary
ZeRO-3 sharding preserves memory but requires matrix-wide communication during every optimizer step.
\textbf{(b)} ZeRO-1 owner placement makes the optimizer local, but full parameters are resident on
every rank. \textbf{(c)} \sys{} makes owner placement a ZeRO-3 layout: backward already leaves the
full gradient on the owner, while forward/backward remain sharded.}
\label{fig:motivation}
\end{figure*}

Figure~\ref{fig:motivation} contrasts the two existing endpoints with the target design point. Under
ordinary ZeRO-3, an exact Muon update cannot be computed independently from equal shards. A
reconstruction path gathers each matrix, runs Newton--Schulz, and redistributes the update; for $N$
2D parameters of $s$ bytes over $W$ data-parallel ranks, this moves roughly
$2Ns\frac{W-1}{W}$ extra bytes per rank per step. Distributed Newton--Schulz avoids full
materialization but instead communicates the coupled matrix products during its iterations. Both
paths place matrix-wide communication after backward, where it cannot overlap layer compute. For
transformers, whose parameter mass is dominated by 2D weights, this recurring optimizer critical path
grows with model size and node count.

The opposite endpoint
removes that optimizer-time reconstruction by assigning whole matrices to owners. Distributed
Shampoo~\cite{distshampoo} and Canzona's data-parallel path~\cite{canzona} assign whole matrices to
owners, so the matrix operation itself is local. The catch is that these systems operate in a
ZeRO-1-style regime with full parameters resident on every rank. They remove optimizer-step
communication by giving up the parameter-memory savings that ZeRO-3 was meant to provide.
\sys{} targets the missing point in Figure~\ref{fig:motivation}: whole-matrix ownership \emph{as} a
ZeRO-3 placement, not as a replacement for ZeRO-3.

\section{Approach: Owner-Shaped ZeRO-3}\label{sec:design}

\sys{} keeps the FSDP training loop but changes the resident shard shape seen after each reshard. For
each 2D parameter $W_i$, a data-parallel rank $o_i$ is chosen as its owner. In the resharded state,
rank $o_i$ stores the whole matrix and every other rank stores an empty shard for that parameter. The
non-2D tensors in the same FSDP unit are packed into a tail role and assigned whole to a tail owner,
where they remain on the AdamW path. Across the data-parallel group there is still only one resident
copy of each parameter; the difference from ordinary ZeRO-3 is that the copy is assigned by tensor
role rather than sliced equally across all ranks. Matrices already fragmented by tensor parallelism are
not owner-placed here; they are skipped by the matrix-owner planner and handled by the surrounding TP
runtime.

This placement makes the optimizer input local. Backward still produces the same per-rank gradients
as ordinary data parallel training, and ZeRO-3 still reduces them into the current local shard shape.
Because the local shard of a 2D matrix is the full matrix on its owner, the reduced Muon input $G_i$
lands on that owner. The owner applies the Newton--Schulz update to the full matrix and updates its
local parameter and Muon state; ranks with empty shards do no optimizer work for that matrix. Tail
owners apply AdamW to their packed non-matrix tensors. Thus the optimizer step contains no matrix
all-gather, broadcast, or redistribute. The next forward pass materializes parameters through FSDP's
usual unshard/reshard lifecycle, now from owner-shaped resident shards.

Owner placement removes post-backward matrix reconstruction, but it makes the resident state uneven.
A usable runtime must solve three resulting problems. First, owner bytes and optimizer work must be
balanced over the whole model even though FSDP2 gathers and prefetches parameters one block at a time.
Second, materialization and gradient reduction must move variable-size owner segments without padding
all ranks to the largest segment or materializing matrices on ranks that do not own them. Third, the
placement must remain valid across FSDP2's transient full-parameter buffers, mixed optimizer state, and
checkpoint save/load. Section~\ref{sec:impl} addresses these problems with a global owner planner,
explicit \texttt{MatrixShard} metadata, owner-segment P2P communication, and placement-aware lifecycle
and state management.

\section{FSDP2 Runtime for Owner Shards}\label{sec:impl}

Ordinary FSDP2 derives communication and state handling from roughly equal, nonempty local shards.
Owner-shaped shards invalidate that assumption: one rank may hold a complete matrix while its peers
hold empty tensors, and each FSDP unit can induce a different per-rank byte distribution. \sys{} makes
ownership explicit at the four runtime boundaries where FSDP2 expects a shard. A global planner emits
one \texttt{MatrixShard} layout per block (\S\ref{sec:impl-shard}); owner-segment communication
materializes and reduces only its nonempty pieces (\S\ref{sec:impl-collectives}); pinned storage keeps
autograd views valid through unshard and reshard (\S\ref{sec:impl-lifecycle}); and the same layout
routes optimizer and checkpoint state (\S\ref{sec:impl-state}). Forward and backward retain FSDP2's
block schedule; only the resident placement and the communication needed to realize it change.

\subsection{\texttt{MatrixShard}: representing ownership}\label{sec:impl-shard}
A standard \texttt{Shard(0)} placement is an equal split of a tensor interval. \sys{} instead records a
rank-indexed placement for each flattened parameter interval. A 2D matrix has exactly one full segment,
stored on its matrix owner, and empty segments elsewhere. The non-2D tail in an FSDP unit is packed
into one whole segment on a tail owner. Figure~\ref{fig:owner-shaped-shards} contrasts this state with
ordinary FSDP2 after reshard: ordinary FSDP2 gives every rank a fraction of every tensor, while
\sys{} gives each tensor role to one owner rank.

\begin{figure}[t]
\vspace{-0.4em}
\centering
\resizebox{0.98\columnwidth}{!}{%
\begin{tikzpicture}[font=\scriptsize, >={Stealth[round]},
  box/.style={draw, rounded corners=1.5pt, minimum height=5.5mm, minimum width=9.5mm, align=center},
  wide/.style={draw, rounded corners=1.5pt, minimum height=6mm, minimum width=18mm, align=center},
  lab/.style={font=\scriptsize\bfseries, align=center},
  flow/.style={->, line width=0.55pt, draw=black!70},
]
\node[lab] at (1.45,1.20) {global owner planner};
\node[font=\scriptsize] at (1.45,0.90) {roles from all blocks};
\foreach \y/\b/\a/\c in {0.45/B1/r0/r2,-0.10/B2/r1/r3,-0.65/B3/r2/r0} {
  \node[font=\scriptsize, anchor=east] at (-0.10,\y) {\b};
  \node[box, fill=green!14, draw=green!55!black] at (0.45,\y) {\a};
  \node[box, fill=green!14, draw=green!55!black] at (1.15,\y) {\c};
  \node[box, fill=orange!11, draw=orange!55!black] at (1.85,\y) {tail};
}
\node[wide, fill=gray!8, draw=black!50] at (1.15,-1.22) {rank load vector};

\draw[flow] (2.35,-0.10) -- (3.10,-0.10);

\node[lab] at (4.65,1.20) {runtime execution};
\node[font=\scriptsize] at (4.65,0.90) {one FSDP block at a time};
\node[wide, fill=blue!8, draw=blue!45!black] (g) at (3.35,0.20) {\begin{tabular}{c}gather\\ block k\end{tabular}};
\node[wide, fill=gray!8, draw=black!55] (c) at (4.65,0.20) {\begin{tabular}{c}compute\\ block k\end{tabular}};
\node[wide, fill=blue!8, draw=blue!45!black] (s) at (5.95,0.20) {\begin{tabular}{c}reshard\\ block k\end{tabular}};
\draw[flow] (g)--(c);
\draw[flow] (c)--(s);
\node[wide, fill=purple!8, draw=purple!45!black] at (4.65,-0.70) {\begin{tabular}{c}prefetch\\ next block\end{tabular}};
\draw[flow, draw=purple!55!black] (3.80,-0.45) -- (5.50,-0.45);
\node[font=\scriptsize, align=center] at (4.65,-1.22) {same per-block schedule as FSDP2};
\end{tikzpicture}%
}
\caption{Owner planning and runtime materialization use different granularities. The planner balances
owners over all wrapped blocks and emits one \texttt{MatrixShard} layout per block; the runtime still
gathers, prefetches, computes, and reshards one block at a time.}
\label{fig:planner-runtime}
\vspace{-0.6em}
\end{figure}
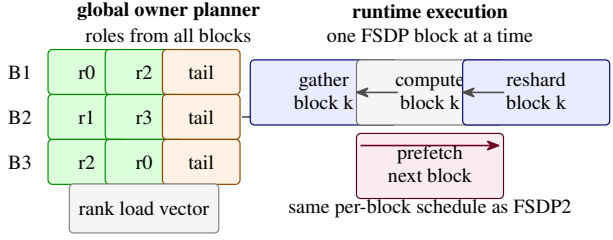

The planner is global, but execution is block-local (Figure~\ref{fig:planner-runtime}). Planning sees
all FSDP blocks selected by the wrap policy. Each block is converted into roles: one role per 2D
matrix and one packed non-2D tail. The greedy load is owner-resident numel, a proxy for resident
parameter/gradient bytes and Muon work; the resource scorer expands a completed plan into per-rank
parameter, gradient, optimizer-state, communication, and workspace estimates. Balance is therefore
decided at model scope, not from one block's local shape. The same owner-byte balance later bounds
the fanout bottleneck in \S\ref{sec:impl-collectives}: if one block assigns too many bytes to one
owner, that rank becomes the source for materialization and the sink for gradient reduction.

Algorithm~\ref{alg:owner-planner} summarizes the owner assignment. \textsc{RoleGreedy} is the
memory-conservative default; \textsc{ScopeGreedy} pools roles from all blocks to improve global balance;
\textsc{CostAware} accepts the scope plan only when the balance gain does not exceed workspace caps.
Each emitted block layout is then checked for one contiguous local segment per parameter, so the hot
path remains a standard FSDP2 block schedule.

\begin{algorithm*}[t]
\footnotesize
\caption{Balance-aware owner planning}
\label{alg:owner-planner}
\vspace{0.2em}
\begin{algorithmic}[1]
\STATE Build roles: each 2D matrix is one role; non-2D tensors form one tail role per block.
\STATE \textsc{RoleGreedy}: per block, assign largest roles to the currently lightest ranks; carry loads forward.
\STATE \textsc{ScopeGreedy}: pool all roles, assign largest roles to lightest ranks, then split layouts by block.
\STATE \textsc{CostAware}: use \textsc{ScopeGreedy} only if balance improves within workspace caps; else use \textsc{RoleGreedy}.
\STATE Validate one contiguous local segment per parameter; return one \texttt{MatrixShard} layout per block.
\end{algorithmic}
\vspace{-0.4em}
\end{algorithm*}

\subsection{Owner-segment communication collectives}\label{sec:impl-collectives}
Once shards are owner-shaped, the equal-size collective path is the wrong abstraction. Padding every
rank to the largest owner segment wastes memory and bandwidth, while all-gathering a matrix to every
rank recreates the optimizer-step reconstruction that owner placement was meant to remove. The logical
operations FSDP2 needs are narrower: materialize a full parameter from its current owner segments, and
reduce the full gradient back into the same owner segments.

\paragraph{Bandwidth model.}
Let \(W\) be the shard span: the data-parallel ranks over which a parameter is sharded, materialized,
and reduced; if world size exceeds \(W\), the remainder is the HSDP replicate dimension. Let \(P_B\)
be the parameter bytes in FSDP block \(B\), and \(O_{B,r}\) the bytes in that block owned by rank \(r\).
We compare three per-block quantities:
\[
\begin{aligned}
\gamma_B
  &= \frac{\max_r O_{B,r}}{P_B/W},\\
C^{\mathrm{FSDP}}_B
  &\approx P_B \frac{W-1}{W},\\
C^{\mathrm{owner}}_B
  &= (W-1)\max_r O_{B,r}
   = \gamma_B P_B \frac{W-1}{W}.
\end{aligned}
\]
Thus a block with one large owner can be bandwidth-poor (\(\gamma_B \approx W\)); when the planner
spreads matrix and tail owners within the block (\(\gamma_B \approx 1\)), the leading bandwidth term
matches ordinary FSDP while the optimizer-step reconstruction is removed.

\sys{} implements these operations as owner-segment collectives. For materialization, ranks with
nonempty segments send only those segments to the ranks that need the full parameter for the current
FSDP unit; in the common matrix case this is an owner fanout. For gradient reduction, each rank
contributes its local full-gradient buffer, but the reduced result is written only to the nonempty owner
segment. The fast path uses deterministic native send/recv schedules; owner-broadcast and uneven torch
collectives are kept as fallbacks. Cross-rank validation checks that every rank derives the same
deterministic schedule and segment sizes before communication begins, turning mismatches into explicit
errors rather than silent NCCL timeouts.

\subsection{Owner-buffer pinning in FSDP2}\label{sec:impl-lifecycle}

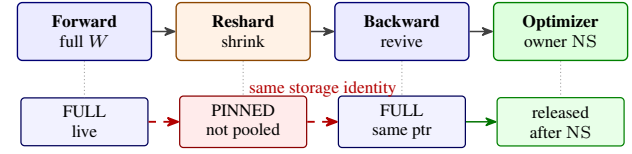
\begin{figure}[H]
\centering
\resizebox{0.98\columnwidth}{!}{%
\begin{tikzpicture}[font=\scriptsize, >={Stealth[round]},
  phase/.style={draw, rounded corners=2pt, line width=0.55pt, minimum height=8.2mm,
                minimum width=19mm, align=center},
  state/.style={draw, rounded corners=1.5pt, line width=0.5pt, minimum height=6.0mm,
                minimum width=18mm, align=center},
  flow/.style={->, line width=0.6pt, draw=black!75},
  guide/.style={densely dotted, line width=0.45pt, draw=black!35},
  hold/.style={->, dashed, line width=0.75pt, draw=red!70!black},
]
\node[phase, fill=blue!8, draw=blue!45!black]    (f)  at (0,0)    {\textbf{Forward}\\full $W$};
\node[phase, fill=orange!10, draw=orange!55!black] (r)  at (2.25,0) {\textbf{Reshard}\\shrink};
\node[phase, fill=blue!8, draw=blue!45!black]    (bw) at (4.50,0) {\textbf{Backward}\\revive};
\node[phase, fill=green!13, draw=green!50!black] (o)  at (6.75,0) {\textbf{Optimizer}\\owner $\NS$};

\draw[flow] (f) -- (r);
\draw[flow] (r) -- (bw);
\draw[flow] (bw) -- (o);

\node[state, fill=blue!5, draw=blue!35!black]       (sf) at (0,-1.28)    {FULL\\live};
\node[state, fill=red!8, draw=red!55!black]         (sr) at (2.25,-1.28) {PINNED\\not pooled};
\node[state, fill=blue!5, draw=blue!35!black]       (sb) at (4.50,-1.28) {FULL\\same ptr};
\node[state, fill=green!9, draw=green!50!black]     (so) at (6.75,-1.28) {released\\after $\NS$};

\draw[guide] (f.south) -- (sf.north);
\draw[guide] (r.south) -- (sr.north);
\draw[guide] (bw.south) -- (sb.north);
\draw[guide] (o.south) -- (so.north);

\draw[hold] (sf.east) -- (sr.west);
\draw[hold] (sr.east) -- (sb.west);
\node[font=\scriptsize, text=red!70!black] at (3.38,-0.82) {same storage identity};
\draw[flow, draw=green!45!black] (sb.east) -- (so.west);
\end{tikzpicture}%
}
\caption{Owner-buffer pinning within FSDP2's unshard/reshard lifecycle. Forward materializes the full
matrix and autograd may save views into that storage. Reshard shrinks the buffer but pins the same
allocation instead of returning it to the pool; backward revives the full matrix in that same storage,
and the owner releases it only after Newton--Schulz.}
\label{fig:lifecycle}
\end{figure}

FSDP2 unshards before forward, reshards after forward, unshards again before backward, and reduces
gradients after backward. \sys{} preserves that lifecycle, but owner-shaped shards expose a
storage-level hazard: autograd saves views into the transient full-parameter buffer. Post-forward
reshard may shrink that buffer in place, but it cannot return the storage to the shared pool. Before
backward, the unit resizes and refills the same storage, so saved views are revived in place
(Figure~\ref{fig:lifecycle}).

The implementation adds a pinned state for shrunk buffers that are pending backward. A central guard
releases storage only after the owning unit finishes backward and optimizer-ready local shards have
been restored. This lets owner-shaped shards coexist with FSDP2's materialize-and-reshard discipline
without \texttt{setStorage ... size 0} errors or rank divergence from reviving a reused buffer.

\subsection{Optimizer and checkpoint state}\label{sec:impl-state}
After backward, every FSDP unit is returned to owner-shaped local shards. Optimizer construction then
uses \texttt{MatrixShard} metadata, not tensor shape alone: a Muon parameter is visible to the matrix
optimizer only on the rank that owns the full 2D shard; ranks with empty shards skip it; and tail-owner
shards are routed to AdamW. Since gradient reduction has already deposited each full 2D gradient on its
owner, Newton--Schulz runs without optimizer-step collectives.

Checkpointing follows the same placement. DCP save writes each local shard as tensor payload plus a
layout sidecar containing the \texttt{MatrixShard} placement, per-rank segments, and replica metadata.
Same-layout load validates the recorded placement before copying payloads into local shards. Reshard
load reconstructs each parameter on a canonical $[0,\text{numel})$ axis from the saved segments, then
repacks it into the target owner assignment and world size. Owner-Muon state and tail-owner AdamW state
use the same path, so changing world size or owner plan does not require optimizer-specific checkpoint
logic.

\begin{figure}[H]
\centering
\resizebox{0.96\columnwidth}{!}{%
\begin{tikzpicture}[
  font=\scriptsize,
  cell/.style={draw, rounded corners=1pt, minimum width=7.2mm, minimum height=4.6mm, align=center},
  head/.style={align=center, font=\scriptsize\bfseries},
  group/.style={align=center, font=\scriptsize\bfseries},
]
\def\xr{0.0}
\def\xa{0.95}\def\xb{1.72}\def\xt{2.49}
\def\ya{3.55}\def\yb{4.32}\def\yt{5.09}
\node[group] at (1.72,0.55) {(a) ordinary FSDP2};
\node[group] at (4.32,0.55) {(b) \sys{}};
\node[head] at (\xa,0.0) {$W_1$};
\node[head] at (\xb,0.0) {$W_2$};
\node[head] at (\xt,0.0) {tail};
\node[head] at (\ya,0.0) {$W_1$};
\node[head] at (\yb,0.0) {$W_2$};
\node[head] at (\yt,0.0) {tail};
\foreach \r/\y in {0/-0.55,1/-1.15,2/-1.75,3/-2.35} {
  \node[anchor=east] at (\xr,\y) {rank \r};
  \node[cell, fill=blue!8] at (\xa,\y) {$1/4$};
  \node[cell, fill=blue!8] at (\xb,\y) {$1/4$};
  \node[cell, fill=blue!8] at (\xt,\y) {$1/4$};
}
\node[cell, fill=gray!8] at (\ya,-0.55) {$\emptyset$};
\node[cell, fill=gray!8] at (\yb,-0.55) {$\emptyset$};
\node[cell, fill=green!18, draw=green!55!black] at (\yt,-0.55) {full};
\node[cell, fill=green!18, draw=green!55!black] at (\ya,-1.15) {full};
\node[cell, fill=gray!8] at (\yb,-1.15) {$\emptyset$};
\node[cell, fill=gray!8] at (\yt,-1.15) {$\emptyset$};
\node[cell, fill=gray!8] at (\ya,-1.75) {$\emptyset$};
\node[cell, fill=gray!8] at (\yb,-1.75) {$\emptyset$};
\node[cell, fill=gray!8] at (\yt,-1.75) {$\emptyset$};
\node[cell, fill=gray!8] at (\ya,-2.35) {$\emptyset$};
\node[cell, fill=green!18, draw=green!55!black] at (\yb,-2.35) {full};
\node[cell, fill=gray!8] at (\yt,-2.35) {$\emptyset$};
\end{tikzpicture}%
}
\caption{Local shards after reshard. Ordinary FSDP2 slices every tensor across ranks; \sys{} keeps
each 2D matrix and the non-2D tail as whole owner roles, with empty shards on the other ranks.}
\label{fig:owner-shaped-shards}
\end{figure}

\section{Evaluation}\label{sec:eval}
We evaluate on a multi-node cluster (Sec.~\ref{sec:eval-setup}). The three headline results: the
optimizer step is communication-free and its speedup over the strongest FSDP2-Muon baseline
\emph{grows with node count} (Sec.~\ref{sec:eval-scaling}); correctness matches DDP Muon
(fp32 tolerance; bf16 printed loss) (Sec.~\ref{sec:eval-conv}); and \sys{} holds ZeRO-3 per-rank memory, far below the
ZeRO-1 owner placement baseline (Sec.~\ref{sec:eval-mem}).

\subsection{Setup}\label{sec:eval-setup}
\textbf{Hardware/software.} All GPU experiments run on 8 nodes $\times$ 8 NVIDIA A100-SXM4-80GB
GPUs (64 GPUs) with NCCL over InfiniBand, using PyTorch FSDP2 \texttt{fully\_shard}, DTensor, and
\texttt{torch.optim.Muon}. \textbf{Models/data.} Latency and memory use bf16 Qwen-style
decoder-transformer shapes
(hidden 4096, intermediate 16384, 32 heads, sequence 4096) with activation checkpointing;
the weak-scaling sweep uses 16/32/64/128 layers for shard spans 8/16/32/64; here shard span equals
world size, so the HSDP replicate dimension is one. Convergence uses a
12-layer decoder-style LM (hidden 768, intermediate 3072, 12 heads, sequence 1024) on a flat uint16
WikiText stream with GPT-2 vocab 50{,}257. The embedding and untied LM head are 2D Muon weights;
LayerNorm and other non-2D tensors use AdamW. Each convergence step is batch 8 per rank on 64 ranks
(524{,}288 tokens/step), for 10{,}000 steps (5.24B tokens), fixed seed 0 and rank-local data streams.
\textbf{Optimizers/timing.} Muon uses lr \(10^{-3}\), weight decay \(10^{-2}\), momentum 0.5,
2 NS steps, and \texttt{match\_rms\_adamw}; AdamW uses lr \(10^{-3}\), weight decay \(10^{-2}\),
betas (0.9, 0.999), eps \(10^{-8}\), with no schedule. Latency uses 3 warmup and 10 measured steps;
we synchronize after \texttt{zero\_grad}, forward, backward, and optimizer step, then report max-rank
phase time. Memory is max-rank peak CUDA allocation after warmup (not reserved cache): it includes
bf16 parameter shards and full-parameter buffers, gradients, optimizer state, \sys{} communication
workspaces, and activations from the fixed checkpointed run; none of the modes uses fp32 master
weights. \textbf{Baselines:} \emph{stock FSDP2-Muon} uses PyTorch
\texttt{torch.optim.Muon} over FSDP2-sharded DTensors; \emph{gather-once FSDP2-Muon} gathers each
matrix gradient once, then runs local NS; \sys{}~= owner-Muon.

\begin{table*}[!t]
\centering\small
\caption{Per-phase latency (ms/step) for a 16-layer, 3.2B-parameter model on one node
(8$\times$A100, shard span 8). \sys{} removes optimizer-step communication; \emph{gather-once FSDP2-Muon} is
slower than stock here, so the headline is against the faster \emph{stock} baseline. The gap grows
with scale (Table~\ref{tab:scaling}).}
\label{tab:speedup}
\begin{tabular}{lrrrr}
\toprule
 & opt-step & fwd & bwd & total \\
\midrule
stock FSDP2-Muon (\texttt{torch.optim.Muon}, NS on DTensor) & 367 & 163 & 460 & 991 \\
gather-once FSDP2-Muon (full matrix $+$ local NS) & 1177 & 164 & 466 & 1808 \\
\sys{} (owner-Muon) & \textbf{87} & 175 & 462 & \textbf{725} \\
\midrule
speedup vs stock (strongest baseline) & 4.2$\times$ & $\approx$1 & $\approx$1 & 1.37$\times$ \\
\bottomrule
\end{tabular}
\end{table*}

\subsection{Optimizer-step and end-to-end speedup}
Table~\ref{tab:speedup} reports one-node latency for a 16-layer, 3.2B-parameter transformer
from the weak-scaling sweep (shard span 8). \sys{}'s optimizer step is
\emph{communication-free}---$87$\,ms, versus $367$\,ms for the strongest correct baseline (stock
FSDP2-Muon, which runs a correct distributed Newton--Schulz on the sharded DTensor): a $4.2\times$
optimizer-step reduction. Forward and backward stay close to the baseline: forward is $175$\,ms versus
$163$\,ms ($1.07\times$), and backward is $462$\,ms versus $460$\,ms. This is the regime predicted by
the \(\gamma_B\) model in \S\ref{sec:impl-collectives}: owner fanout is not universally better than a
ring all-gather, but with balanced block owners it does not dominate the materialization path while the
optimizer-step reconstruction disappears. \emph{Gather-once FSDP2-Muon} is in fact
\emph{slower} than stock here---materializing each full matrix on every rank and re-running NS
redundantly costs more than PyTorch's sharded NS---so we headline against the faster stock baseline.
This single-node $4.2\times$ is the small end of the story: Sec.~\ref{sec:eval-scaling} shows the
optimizer-step gap widening to $54\times$ at 8 nodes as the baseline's per-step matrix communication
crosses the inter-node fabric while \sys{}'s step stays local.

\subsection{Convergence and numerical checks}\label{sec:eval-conv}
\sys{} should change data movement, not the optimizer being computed. The semantic reference is
therefore full-matrix Muon, implemented as DDP Muon: every rank holds full parameters, gradients are
averaged over data-parallel ranks, and Newton--Schulz runs on each complete 2D gradient. FSDP2-Muon
remains the systems baseline for speed and memory, but DDP Muon is the cleaner correctness oracle
because it avoids FSDP-specific reconstruction and bucket-order effects.

We check equivalence at two levels. First, a deterministic float32 harness disables TF32, fixes the
decoder-LM batches, and uses distinct rank-local inputs. It compares per-step losses over a short
training trajectory, final logits after the resulting updates, and reconstructed full gradients for
every parameter at the final backward pass. Agreement within tolerance directly tests both the
gradient delivered to each owner and the resulting optimizer trajectory.

\begin{figure}[t]
\centering
\includegraphics[width=\columnwidth,trim=0 24pt 0 0,clip]{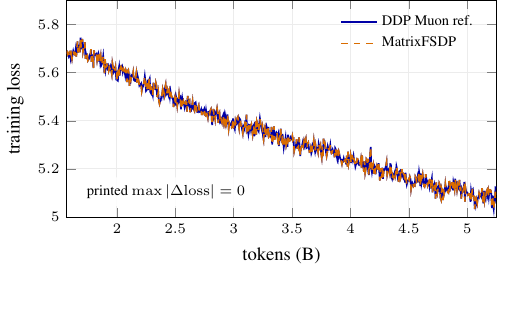}
\caption{Real-data convergence on WikiText after step 3000. \sys{} overlays a DDP Muon reference
(bf16, identical initialization and token stream); the full 10{,}000-step log records max printed
\(|\Delta\mathrm{loss}|=0\).}
\label{fig:loss_curve}
\vspace{-0.4em}
\end{figure}

\enlargethispage{\baselineskip}
Second, both modes train the same 12-layer decoder-only LM on the same WikiText token stream for
10{,}000 bf16 steps (5.24B tokens), starting from identical initialization. The DDP reference uses
deterministic per-parameter all-reduces rather than buckets, removing reduction-order noise.
Figure~\ref{fig:loss_curve} shows steps 3000--10{,}000; losses match at printed precision, not
bitwise identity for arbitrary bf16 reduction orders.

\subsection{Memory}\label{sec:eval-mem}
\begin{figure}[t]
\centering
\includegraphics[width=\columnwidth,trim=0 22pt 0 0,clip]{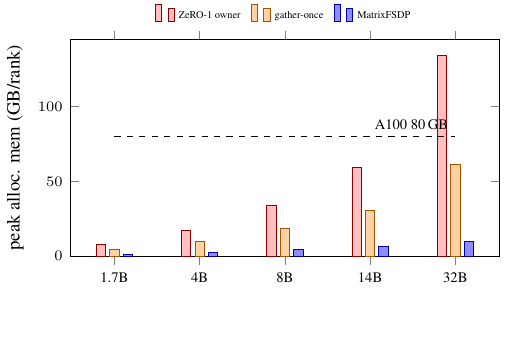}
\caption{Per-rank peak allocation at shard span 64. \sys{} uses 10\,GB/rank at 32B and
$3$--$7\times$ less memory than gather-once FSDP2-Muon. ZeRO-1 owner placement exceeds 80\,GB at
$\ge$14B; values above 12.9B are projected.}
\label{fig:memory_cliff}
\vspace{-0.4em}
\end{figure}

Owner placement only solves the optimizer communication problem if it does not also give up ZeRO-3
memory. Table~\ref{tab:memory} compares \sys{} with ZeRO-1 owner placement, where full parameters and
gradients are resident on every rank. For the 4.3B model, ZeRO-1 owner placement
uses $18.5$\,GB/rank; \sys{} keeps the model in owner-shaped ZeRO-3 shards at $2.7$\,GB/rank, while
also reducing the optimizer step from $677$\,ms to $19$\,ms. The model-size sweep in
Figure~\ref{fig:memory_cliff} shows that this is not a constant offset: ZeRO-1 owner placement reaches
$54.5$\,GB/rank at 12.9B and crosses the 80\,GB A100 limit at $\ge$14B parameters, whereas \sys{} uses
$6.3$\,GB/rank at 14B and $10$\,GB/rank at 32B. Relative to stock FSDP2-Muon, \sys{} uses
$0.6$\,GB/rank more peak allocation at this point ($2.7$ versus $2.1$\,GB, a $29\%$ premium) in
exchange for removing optimizer-step matrix communication. As defined in \S\ref{sec:eval-setup}, this
peak includes parameters, full-parameter buffers, gradients, optimizer state, communication
workspaces, and checkpointed activations; neither mode keeps fp32 master weights.

\begin{table*}[t]
\centering\small
\caption{Per-rank peak allocated memory and optimizer-step latency, 4.3B model sharded over 64 GPUs.
ZeRO-1 owner placement keeps full parameters resident; \sys{} keeps ZeRO-3 memory \emph{and} the
fastest step. At 12.9B the memory gap to ZeRO-1 owner placement grows to $15\times$ ($54.5$ vs
$3.6$\,GB/rank).}
\label{tab:memory}
\vspace{0.25em}
\begin{tabular}{lrr}
\toprule
 & opt-step & peak alloc. mem \\
 & (ms) & (GB/rank) \\
\midrule
ZeRO-1 owner placement & 677 & 18.5 \\
stock FSDP2-Muon & 651 & 2.1 \\
\sys{} (ZeRO-3 owner-Muon) & \textbf{19} & \textbf{2.7} \\
\bottomrule
\end{tabular}
\vspace{-0.45em}
\end{table*}

\subsection{Scaling}\label{sec:eval-scaling}
\sys{}'s advantage is fundamentally a \emph{scaling} property (Table~\ref{tab:scaling}). This weak
scaling grows depth with shard span; fixed-model strong scaling can reduce owner-role density and raise
\(\gamma_B\) (\S\ref{sec:impl-collectives}). The baseline's
optimizer step grows from $367$\,ms on 1 node to $5064$\,ms on 8 nodes as its per-step matrix
communication crosses the inter-node fabric; \sys{}'s step stays flat at $\sim$90\,ms because it issues
no optimizer-step collectives. The optimizer-step speedup therefore grows $4.2\times \to 54.6\times$
over $1\to8$ nodes, and end-to-end step time $1.37\times \to 2.15\times$ (peaking at 4 nodes).
The end-to-end ratio is not monotone because the non-optimizer phases also scale with depth,
materialization, gradient reduction, and overlap; once \sys{}'s optimizer step is flat, these phases
set the residual denominator.
Forward, backward, and per-rank peak allocation stay near FSDP2 across the sweep because owner
materialization and gradient reduction remain block-local and are issued on the communication stream,
preserving the same overlap opportunities as FSDP2.

\begin{figure}[t]
\centering
\includegraphics[width=\columnwidth]{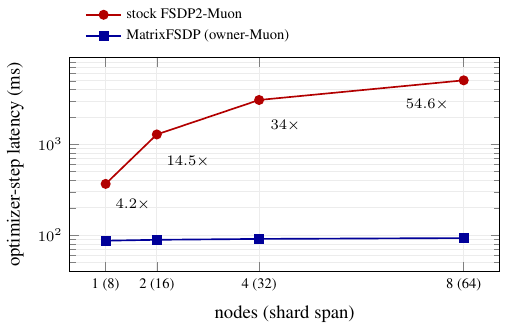}
\caption{Optimizer-step latency vs node count (weak scaling, depth scaled with shard span, 8--64 GPUs,
InfiniBand; log scale). The strongest correct baseline (stock FSDP2-Muon) grows $367\to5064$\,ms as its
per-step Newton--Schulz matrix communication crosses the inter-node fabric, while \sys{}'s
communication-free step stays flat ($\sim$90\,ms). The optimizer-step speedup (labels) therefore
\emph{grows} with scale, $4.2\times\to54.6\times$ (end-to-end speedup reaches $2.15\times$):
\sys{}'s advantage is a scaling property, not a fixed constant.}
\label{fig:scaling}
\end{figure}

\begin{table*}[t]
\centering\small
\caption{Weak-scaling sweep, 1--8 nodes (shard span 8--64; world size equals span; depth scaled with
span; hidden 4096).
Optimizer-step latency (ms) for the strongest baseline (stock FSDP2-Muon) vs \sys{}, plus end-to-end
step time. The optimizer-step gap grows with node count because the baseline's matrix communication
crosses the inter-node fabric while \sys{}'s step is local; forward/backward/peak allocation stay at FSDP2
parity throughout.}
\label{tab:scaling}
\vspace{0.25em}
\resizebox{0.92\textwidth}{!}{%
\setlength{\tabcolsep}{4pt}
\begin{tabular}{@{}llrrrrrr@{}}
\toprule
nodes (span) & model & stock opt & \sys{} opt & opt speedup & stock total & \sys{} total & e2e speedup \\
\midrule
1 (8)  & 16L / 3.2B  &  367 & 87 & 4.2$\times$  &  991 &  725 & 1.37$\times$ \\
2 (16) & 32L / 6.4B  & 1285 & 89 & 14.5$\times$ & 2612 & 1489 & 1.75$\times$ \\
4 (32) & 64L / 12.9B & 3079 & 91 & 34.0$\times$ & 5788 & 2689 & 2.15$\times$ \\
8 (64) & 128L / 25.8B & 5064 & 93 & 54.6$\times$ & 10004 & 4989 & 2.01$\times$ \\
\bottomrule
\end{tabular}
}
\vspace{-0.35em}
\end{table*}

\subsection{Model-size scaling}\label{sec:eval-size}
The communication-free optimizer also holds across model size. Unlike Table~\ref{tab:scaling},
Table~\ref{tab:sizesweep} fixes shard span at 64 and varies model size, measuring larger matrices at
the same sharding degree. \sys{}'s optimizer step is $6$--$160$\,ms versus
$1.0$--$6.2$\,s for stock FSDP2-Muon (which runs distributed Newton--Schulz collectives across all 64
ranks every step):
a $31$--$159\times$ optimizer-step and $1.85$--$3.3\times$ end-to-end speedup, with forward and backward
at parity. These ratios are not expected to be monotone: model shape and the forward/backward share
change across rows, while the removed optimizer communication is only one part of total step time.
Per-rank peak allocation stays at ZeRO-3 scale---$1.4$--$10.0$\,GB/rank---and is $3$--$7\times$ below
gather-once FSDP2-Muon, which materializes full matrices.

\begin{table*}[t]
\centering\small
\caption{Model-size sweep at fixed 64-GPU ZeRO-3 sharding (shard span 64, replicate dimension one),
1.7B--32B. Optimizer-step/end-to-end latency vs stock FSDP2-Muon, and peak per-rank allocated memory
(\sys{} vs gather-once FSDP2-Muon).}
\label{tab:sizesweep}
\setlength{\tabcolsep}{4pt}
\begin{tabular}{@{}lrrrrrr@{}}
\toprule
model & stock opt (ms) & \sys{} opt (ms) & opt speedup & e2e speedup & \sys{} mem (GB) & gather-once mem (GB) \\
\midrule
1.7B & 1016 &   6.4 & 159$\times$ & 3.3$\times$  &  1.4 &  4.4 \\
4B   & 1235 &  14.8 &  83$\times$ & 2.4$\times$  &  2.6 &  9.7 \\
8B   & 1583 &  34   &  47$\times$ & 2.0$\times$  &  4.3 & 18.4 \\
14B  & 2286 &  73   &  31$\times$ & 1.85$\times$ &  6.3 & 30.7 \\
32B  & 6219 & 160   &  39$\times$ & 2.2$\times$  & 10.0 & 61.6 \\
\bottomrule
\end{tabular}
\end{table*}

\paragraph{Load balance.} The global owner planner keeps the assignment even: across the sweep the
per-rank optimizer-compute imbalance (max/avg step time) is $1.26$--$1.39$ and resident-parameter-memory
imbalance is $1.18$--$1.21$. This is competitive with Canzona's $\alpha$-balanced static partition, which
reports a $1.43\times$ load-balance ratio (improved from $3.24\times$), even though \sys{} must also keep
execution block-local for FSDP2 prefetch and overlap.

\subsection{Generality beyond Muon}
The owner-placement runtime is optimizer-agnostic: each 2D matrix is whole on its owner, so any
\emph{single-device} matrix optimizer runs there with no gather and no distributed variant. We verify this
by swapping owner-Muon for Shampoo (full $L$/$R$ preconditioners with inverse fourth root) and SOAP (Adam
in the Shampoo eigenbasis) while keeping the same placement, collectives, routing, and checkpointing.
Against a \emph{gathered} reference for each optimizer, 64-GPU float32 owner-placed loss trajectories
match to $\le$\!$4\mathrm{e}{-5}$ relative for Shampoo and $\le$\!$1.3\mathrm{e}{-4}$ for SOAP. The
zero-optimizer-step-communication property is therefore inherited structurally: heavy preconditioner
compute runs once per matrix on its owner instead of redundantly across ranks.

\subsection{Ablations}
Table~\ref{tab:ablation-owner-collective} separates two mechanisms in the owner-Muon path: which
rank owns each matrix, and how non-owner ranks materialize owner-shaped matrices during
forward/backward. All rows use the same 64-layer, 12.9B-parameter model; the intended comparisons are
\emph{within a row}: for a fixed span, change the owner policy or collective path. Among the three
custom-collective owner policies, total step time differs by at most $3\%$. The balance column reports
the per-rank optimizer-compute max/avg imbalance for the role-, scope-, and cost-aware planners
(\texttt{r/s/c}); values closer to $1$ mean a flatter owner assignment. \texttt{scope\_greedy} is
flattest, while \texttt{role\_greedy} is the memory-conservative default.

The collective implementation is not second-order. In each row, replacing the owner-segment custom
collective with a matrix all-gather keeps the optimizer step local, but makes the forward/backward
materialization path much more expensive. At the largest point, end-to-end time rises from $2.51$\,s
to $18.08$\,s ($7.2\times$), with forward $9.4\times$ and backward $6.6\times$ slower. Thus the owner
placement alone is insufficient; custom P2P owner collectives are the mechanism that makes the layout
usable inside FSDP2's normal materialize/reshard loop. In the terms of \S\ref{sec:impl-collectives},
the custom path moves the balanced owner segments, while the all-gather variant retreats to moving full
matrices through the materialization path. We do not report the raw CUDA peak-allocation field from
this diagnostic run as a memory result: it includes transient collective workspaces and allocator
state, whereas Sec.~\ref{sec:eval-mem} measures the resident ZeRO-3 memory question.

\begin{table*}[t]
\centering\small
\caption{Owner-policy and collective ablation on the same 64-layer, 12.9B-parameter model. Times are
end-to-end ms/step; role/scope/cost use custom owner-segment collectives, while all-gather replaces
that collective under role-greedy. Balance is optimizer max/avg for role/scope/cost
(\texttt{r/s/c}; lower is better); slowdown is all-gather divided by role.}
\label{tab:ablation-owner-collective}
\setlength{\tabcolsep}{4pt}
\begin{tabular}{@{}lrrrrrl@{}}
\toprule
nodes (span) & role (ms) & scope (ms) & cost (ms) & all-gather (ms) & slowdown & balance (r/s/c) \\
\midrule
1 (8)  & 2833 & 2842 & \textbf{2830} &  3114 & 1.10$\times$ & 1.07/1.02/1.07 \\
2 (16) & 2736 & 2809 & \textbf{2729} &  5817 & 2.13$\times$ & 1.05/1.02/1.05 \\
4 (32) & 2565 & 2572 & \textbf{2560} &  9786 & 3.82$\times$ & 1.12/1.03/1.11 \\
8 (64) & 2511 & \textbf{2486} & 2508 & 18079 & 7.20$\times$ & 1.27/1.04/1.27 \\
\bottomrule
\end{tabular}
\end{table*}

\section{Related Work}\label{sec:related}

\paragraph{Matrix-structured optimizers.} Coordinate-wise adaptive optimizers such as
Adam/AdamW~\cite{adam,adamw} remain the default path for non-matrix parameters, and memory-saving
variants such as Adafactor~\cite{adafactor} factor optimizer state for large matrices. Muon~\cite{muon}
orthogonalizes the momentum via Newton--Schulz and has shown strong token-efficiency, validated at
frontier scale~\cite{moonshot,essentialai}. Shampoo~\cite{shampoo,anil2020} and SOAP~\cite{soap} are
related whole-matrix preconditioned optimizers; NorMuon~\cite{normuon} refines Muon with neuron-wise
normalization. Unlike coordinate-wise or factored-state methods, these optimizers require a
\emph{whole-matrix} per-weight update, the tension this paper targets.

\paragraph{Distributed Muon under data parallelism.} The dominant approach reconstructs each matrix per
step. Moonshot's distributed Muon~\cite{moonshot} uses ZeRO-1 and, per owned partition, all-gathers the
full gradient, runs NS, and discards non-owned shards. In the FSDP2/ZeRO-3 world the community
catalog~\cite{mainhorse,torchtitan} enumerates all-gather-then-NS, per-rank-NS-plus-all-reduce, and
gather-to-owner-then-broadcast variants; NorMuon's FSDP2 implementation~\cite{normuon} spreads the
orthogonalization compute across ranks, and Megatron Core~\cite{megatronmuon} ships duplicated,
distributed, and (approximate) blockwise NS modes. Apart from that approximate mode, all keep equal
sharding and communicate matrix-wide state every step; our owner placement removes that per-step
reconstruction.

\vspace{-0.7em}
\paragraph{Owner placement and logical/physical decoupling.} PyTorch Distributed
Shampoo~\cite{distshampoo} assigns whole parameter blocks to DDP ranks via DTensor, computes the
preconditioner on the owner, and all-gathers the \emph{search direction}---owner placement, but under
replicated parameters. Concurrently, Canzona~\cite{canzona} decouples logical optimizer ownership from
physical placement for Muon/Shampoo/SOAP: its data-parallel path uses ZeRO-1 owners, and its
tensor-parallel path reconstructs fragments asynchronously within the intra-node domain. \sys{} is
orthogonal: it realizes owner placement \emph{inside ZeRO-3 parameter sharding}. This keeps ZeRO-3
per-rank memory but requires runtime pieces ZeRO-1 owner systems avoid: \texttt{MatrixShard}
placement, block-local owner collectives, owner-buffer pinning, and owner-shard checkpointing.
Tensor-parallel coverage, including Canzona-style TP fragment reconstruction, is future/orthogonal.

\vspace{-0.7em}
\paragraph{Concurrent systems.} Two concurrent systems corroborate this design point.
DMuon~\cite{dmuon} independently arrives at owner-centric execution, composed as a drop-in module
beside FSDP2: managed matrices leave the FSDP lifecycle for DMuon's own hooks, owner buffers, and
broadcast/reduce streams, so a training step runs two independently scheduled parameter lifecycles;
its complementary contributions are a profiled MILP owner assignment and a Gram-space NS kernel
stack, and it evaluates against all-gather-then-NS. \sys{} instead expresses ownership as the ZeRO-3
shard layout inside the single FSDP2 lifecycle---one scheduler orders, prefetches, and
cross-rank-validates every collective---and adds the \(\gamma_B\) bandwidth model, pinned-buffer
lifecycle correctness, owner-shard checkpoint resharding, and Shampoo/SOAP generality, benchmarking
against the stronger stock sharded-NS baseline.
veScale-FSDP~\cite{vescalefsdp} generalizes FSDP shards to a block-granular \texttt{RaggedShard}
format and supports Muon by redistributing each matrix to a root rank at optimizer time; that path
still moves matrix-wide bytes per step, whereas \sys{}'s backward reduction already deposits the full
gradient at its owner, so the optimizer step issues no collectives.

\paragraph{Algorithmic alternatives.} Dion~\cite{dion} changes the optimizer instead: a
low-rank/amortized power iteration that keeps matrices sharded and synchronizes only low-rank
factors---communication-friendly but not numerically equal to exact Newton--Schulz.
SignMuon~\cite{signmuon} and MuonBP~\cite{muonbp} likewise trade exactness for communication, via
1-bit majority-vote aggregation and block-periodic orthogonalization respectively. Our approach keeps
the optimizer exact and changes the layout.

\vspace{-0.8em}
\paragraph{Distributed training systems.} Synchronous data-parallel runtimes such as Horovod and
PyTorch DDP~\cite{horovod,pytorchddp} make replicated-parameter training fast through efficient
all-reduce, bucketing, and communication/computation overlap. Tensor, pipeline, and compiler-planned
parallel systems~\cite{gpipe,pipedream,megatronlm,megatronlm2,meshtensorflow,gshard,gspmd} split
computation or tensors across devices and are complementary to data-parallel sharding. ZeRO and
FSDP/FSDP2~\cite{zero,zerooffload,zeroinfinity,fsdp,torchtitan} provide the sharded-parameter substrate;
we extend their equal-sharding placement with whole-parameter owner roles for matrices and non-matrix
tails, expressed as a custom DTensor placement.

\section{Conclusion}\label{sec:concl}
\sys{} is useful when sharding makes the optimizer step, rather than the model layers, the expensive
part of training. It keeps each matrix whole only at its assigned owner, so Muon-style updates run
locally while forward, backward, and checkpointing still follow the ZeRO-3 discipline.

This is the strongest fit for multi-node ZeRO-3 runs with large shard spans, where reconstructing
matrices at every optimizer step would cross the inter-node fabric. The gain is smaller on single-node
runs, when high gradient accumulation amortizes one optimizer step over many forward/backward passes,
or under fixed-model strong scaling where \(W\) grows far beyond the number of sizable owner roles in
a block. In that last regime, \(\gamma_B\) can grow and owner fanout can become the next bottleneck,
making hierarchical owner-fanout a natural extension.

\bibliographystyle{mlsys2024}
\bibliography{matrixfsdp}

\end{document}